# Which kind of research papers influence policymaking

Pablo Dorta-González

Institute of Tourism and Sustainable Economic Development (TIDES), Campus de Tafira, University of Las Palmas de Gran Canaria, 35017 Las Palmas de Gran Canaria, Spain ORCID: http://orcid.org/0000-0003-0494-2903

pablo.dorta@ulpgc.es

**Abstract**

This study examines the use of evidence in policymaking by analysing a range of journal and article attributes, as well as online engagement metrics. It employs a large-scale citation analysis of nearly 150,000 articles covering diverse policy topics. The findings highlight that scholarly citations exert the strongest positive influence on policy citations. Articles from journals with a higher citation impact and larger Mendeley readership are cited more frequently in policy documents. Other online engagements, such as news and blog mentions, also boost policy citations, while mentions on social media X have a negative effect. The finding that highly cited and widely read papers are also frequently referenced in policy documents likely reflects the perception among policymakers that such research is more trustworthy. In contrast, papers that derive their influence primarily from social media tend to be cited less often in policy contexts.

## 1. Introduction

Scientific research has value not only in the academic sphere, as typically measured by citation metrics, but also in its broader societal impact. Societal impact encompasses the impact of research across different sectors. In a study by Wilsdon et al. (2015), societal impact was defined to include its influence on education, society, culture, and the economy. In this context, the altmetrics, as advocated by the NISO Alternative Assessment Metrics (NISO 2016), provide a quantitative approach to measuring the wider impact of publications.

The evolution of digital scholarly communication has led to a significant shift in how the societal impact of scholarly research is assessed. This shift has fostered a more comprehensive approach that encompasses a wider range of scholarly publications and innovative communication methods (as evidenced by surveys conducted by Bornmann 2013; de Rijcke et al. 2016; Bornmann and Haunschild 2019). The implementation of the Research Excellence Framework (REF 2021) in the UK is a notable example of the assessment of research quality in higher education institutions. Within this framework, the assessment of impact beyond the scientific domain is of considerable importance, accounting for 25% of the overall assessment. This includes measuring the impact of research on public policy, services, the economy, society, culture, health, the environment, and overall quality of life (see Khazragui and Hudson 2015).

The research community is increasingly recognizing the need to reconsider altmetrics concerning impact (Spaapen & van Drooge 2011; Joly et al. 2015; Morton 2015). Rather than functioning as direct indicators of impact, recent studies suggest that altmetrics are better understood as tools for analysing how research engages with society and how knowledge circulates beyond academic boundaries (Haustein et al. 2016; Ravenscroft et al. 2017). This view is further reinforced by scholars advocating for a rethinking of altmetrics (Robinson-García et al. 2018; Wouters et al. 2019), with ongoing research continuing to refine this evolving framework (Costas et al. 2021; Alperin et al. 2023; González-Betancor & Dorta-González 2023).

The inclusion of scientific articles in policy documents serves as a strong indicator of the impact of research on society (Yu et al. 2023). Furthermore, the citation of research in policy documents enhances the credibility of both the referenced authors and the documents themselves, as highlighted by Bornmann et al. (2016), thus providing valuable insights into the symbiotic relationship between academic research and policymaking.

Despite the ongoing pursuit of evidence-based policymaking, there remains a persistent gap between the generation of scientific knowledge and its use in policy formulation. This study aims to address this crucial gap by examining the factors that influence policymakers' referencing of research. By identifying the different citation practices of policymakers and other stakeholders, this study sheds light on the most effective communication channels for narrowing the knowledge gap between science and policy, to improve the translation of scientific knowledge into actionable policy decisions.

Unlike the approach adopted by Dorta-González et al. (2024), which focused only on journals containing the term 'policy' in their titles, this study expanded the database to include all journals categorised as policy-related by the Australian Political Studies Association (APSA) expert group.

This decision aimed to capture a broader sample of policy research, reflecting its interdisciplinary nature. In addition, APSA journal ratings were included alongside two measures of journal impact to provide a comprehensive assessment of the influence of policy research. Moreover, social media mentions were disaggregated by platform and some categories such as patents and videos were introduced, allowing for a more detailed analysis of policy research dissemination and engagement across different channels and audiences.

## 2. Literature review

*Connections between Research and Policy*

According to a study by Willis et al. (2017), citations emerged as one of the top eight influential factors considered by stakeholders when assessing social outcomes in policy papers. Another investigation by Yin et al. (2021) explored the intersection of science and policy during the COVID-19 pandemic, using data from Overton. Their findings showed that a significant proportion of pandemic-related policy papers relied heavily on recent and influential peer-reviewed scientific studies. Also, in the context of COVID-19, Dorta-González (2023) examined the correlation between citations in policy documents and mentions on Twitter and found a significant positive association between both variables.

In addition, a meta-analysis by Abbott et al. (2022) examined the relationship between the quality of reviews and the level of interest from researchers, policymakers, and the media in the early COVID-19 published evidence syntheses. Although only a limited number of reviews were cited in policy documents, the study raised concerns about the influence of review quality on their citation in policy documents.

In another recent investigation, Bornmann et al. (2022) explored the relationship between policy and research on climate change. Their findings showed that articles referenced in climate change policy documents received significantly higher citation counts than those that were not referenced. In addition, their model highlighted the different ways in which scientific research influences policy development in different types of papers.

In contrast, Newson et al. (2018), in their quest to assess the practicality and impact of research, used a retrospective tracing method to assess eighty-six policy documents from New South Wales focused on childhood obesity. Their conclusions suggest that, in this scenario, mentions of cited research in policy documents are insufficient to demonstrate its impact on the policymaking process.

*Data aggregators for policy documents*

The Altmetric Attention Score, developed by Altmetric.com, is a measure of the total attention a research result has received. Each contributing source is given a weight, with policy documents given a weight of 3 (compared to 8 for news, 5 for blogs, 3 for Wikipedia, and 0.25 for Facebook and X/Twitter - posts and reposts).

Previous studies have shown that citations of research in policy documents receive limited coverage in the Altmetric.com dataset, compared to mentions on X/Twitter and Facebook. Bornmann et al. (2016) found that only 1.2% of research papers received at least one citation in policy documents. Haunschild and Bornmann (2017) reported that less than 0.5% of papers in different subject categories were cited at least once in policy-related documents.

According to research conducted by Tattersall and Carroll (2018) at the University of Sheffield, only 1.41% of the 96,550 research outputs tracked by Altmetric.com were cited in policy documents. This low citation rate raises concerns about the effectiveness of multidisciplinary research in influencing policy decisions. In contrast, a recent study by Szomszor and Adie (2022) showed that the new altmetrics database, Overton, provides more comprehensive coverage of policy document citations than Altmetric.com. This highlights the importance of using accurate and comprehensive data sources when assessing the impact of policy research.

To assess the link between multidisciplinary research and its adoption in policy documents, Pinheiro et al. (2021) paired the Overton database with Scopus data. Their results showed a coverage rate of 6.0% for all funded publications in the dataset, suggesting a higher proportion of multidisciplinary research.

Haunschild and Bornmann (2017) identify several factors that contribute to the limited presence of citations in policy documents to research papers: First, the sources Altmetric.com uses for policy documents are limited in scope, leading to incomplete data coverage. Second, as much of the scientific literature is primarily aimed at an academic audience, only a fraction may be directly relevant to policy issues. Thirdly, policy documents may not follow scientific citation conventions, and their authors, who are often not researchers themselves, may only sporadically refer to scientific studies. Finally, there may be barriers and poor communication between policymakers and researchers.

*Reasons for using alternative metrics*

Exploring the motivations behind references to research in policy documents is an understudied area. However, numerous studies have sought to understand the motivations behind engagement on various social platforms such as blogs, Facebook, X/Twitter, and Sina Weibo. These studies have used interview and content analysis methods to uncover these motivations.

Shema et al. (2015) conducted a content analysis of blog posts within the health category of Researchblogging.org and identified broad themes of motivation including discussion, critique, guidance, trigger, extension, self, controversy, data, ethics, and others. Notably, the most cited motivations were guidance, critique, and conversation.

Academics have also explored the reasons for disseminating scholarly work on X/Twitter. Veletsianos (2012) analyzed the tweets of forty-five academics and found that sharing knowledge, resources, and media was the main purpose. Na (2015) conducted a content analysis to explore the motivations behind English tweets referencing academic publications in psychology, identifying discussion as the primary motivator, with a significant subset devoted to describing and interpreting scientific findings. Yu et al. (2017) used content analysis in the Sina Weibo environment and identified four main drivers of scientific engagement: discussion, marketing, triggering, and distribution.

In addition, Syn and Oh (2015) conducted an online survey to investigate the elements that motivate people to share information on Facebook and X/Twitter. Their research identified motivators such as enjoyment, efficacy, learning, self-gain, compassion, empathy, social engagement, community interest, reputation, and reciprocity. Users share information on these platforms with the expectation of feeling involved and connected to online communities.

3. **Methodology**

*Objectives of the study*

This study seeks to address a gap in the current literature by examining the factors influencing citations in policy documents to research papers, an area that has been under-researched. While previous research has explored the motivations behind research citations on social media platforms, there remains a lack of understanding of the determinants of citations in policy documents. To address this gap, we conducted a regression analysis to investigate the potential impact of diverse types of influence and bibliometric variables on policy citations.

*Data*

The Australian Political Studies Association (APSA) provides a ranking of political science journals on its website (APSA 2022). This ranking, updated in 2022, includes 703 journals that are categorized into four levels: A* (top 5%), A (next 15%), B (next 30%), and C (next 50%). The evaluation criteria for these tiers include the quality of the research, the diversity of the audience, and the impact of the journal within the field of political science. Of these journals, 488 were indexed in Scopus at the time of this study.

The tier descriptions by McDonnell and Morgenbesser (2019) define A* journals as the pinnacle of their field or subfield, typically publishing high-quality articles and having low acceptance rates. Tier A journals maintain high standards with notable researchers on their editorial boards, while Tier B includes journals with solid reputations but less prominent international recognition. Tier C includes high-quality, peer-reviewed journals that do not meet the criteria for higher tiers.

In order to compile a comprehensive dataset of policy-relevant journal articles, it was necessary to address limitations in the APSA list, which does not provide International Standard Serial Numbers (ISSNs) for the included journals. To overcome this, we conducted a systematic search for journal titles within the Scopus database. Scopus was used exclusively to extract the journal-level variables, which include the ISSNs, Scimago Journal Rank (SJR), the journals' best quartile classification, and the average number of citations per document over a three-year period. This ensured that all journal-level metrics were derived consistently from a single, reliable source.

Subsequently, the retrieved ISSNs were utilised to cross-reference records in the Altmetric.com database. The Altmetric.com search was constrained to publications within the period 2014–2023. From Altmetric.com, we further extracted a range of article-level metrics and additional variables, resulting in a final dataset encompassing N = 149,557 policy journal articles that are indexed concurrently in both Scopus and Altmetric.com. Data collection was finalised on 5 April 2024.

This methodological approach not only harmonises the extraction of both journal-level and article-level indicators by relying on Scopus and Altmetric.com respectively, but it also enhances the reproducibility and technical rigour of our data assembly process.

It is important to note that previous studies have highlighted significant differences in the coverage of policy document citations between Overton and Altmetric. For instance, a recent study found that Overton identified more citations in policy documents than Altmetric.com (Dorta-González et al. 2024). These discrepancies may stem from the different methodologies and sources each platform employs to track citations in policy documents, which could introduce biases in the assessment of research impact on policymaking.

Concerns about the representativeness of the sample analysed stem from the multidisciplinary or interdisciplinary nature of the field of public policy. This characteristic implies that certain policy studies may be published in journals that are not included in the APSA list of policy journals. Consequently, the absence of such studies is likely to vary unevenly across policy areas.

Nevertheless, there are arguments to support the notion that this sample adequately represents the policy domain.

The classification of access types (gold, hybrid, green, bronze, closed) was derived directly from the paper itself. The Altmetric database captures access types at the document level, allowing nuanced distinctions between papers within the same journal. For example, a journal issue may contain articles with different access typologies, such as paywall access (closed OA), OA provided by the publisher to increase citations and journal impact (bronze OA), or OA facilitated by thematic or institutional repositories (green OA). Each type of access has different implications for readership, visibility, and scholarly impact.

The variables and metrics analysed in this study are described below, both at the journal and article level.

*Journal-level metrics*

*Journal expert rating:* This metric represents the assessment or rating of a journal's overall quality, credibility, and significance within the field of political studies by experts affiliated with the Australian Political Studies Association (APSA). It reflects the collective judgment of experts familiar with the journal's content, editorial policy, and impact within the political studies community.

*Journal prestige score SJR*: The Scimago Journal Rank (SJR) is a widely recognized metric developed by Scimago Lab to measure the prestige and visibility of academic journals. It ranks journals based on the number of citations received by the articles they publish. This assessment considers both the quantity and quality of citations. Citations are weighted according to the influence of the citing journal. Therefore, not all citations carry the same weight.

*Journal best quartile SJR*: This metric ranks journals into quartiles based on their SJR scores, with the top quartile representing journals with the highest prestige and impact. Journals in the top quartile are among the most influential and prestigious in their fields, including political science.

*Journal impact score* (cites per doc 3 years): This metric measures the average number of citations received per document (article) published in a journal over a given period of three years.

*Article age*: This attribute indicates the amount of time that has elapsed since an article was published. It is measured in years and can influence its impact.

*Funding*: This attribute refers to the financial support received to conduct the research presented in an article. Funding sources may include government agencies, private foundations, non-profit organizations, industry sponsors, or academic institutions. Funding can cover various aspects of research, including personnel salaries, equipment, supplies, travel expenses, and publication fees.

*Closed access*, also known as subscription access, refers to the traditional model in which scholarly articles are accessible to readers only through subscription or paywall barriers. Articles are typically only available to subscribers or those affiliated with subscribing institutions, limiting access to a wider audience.

*Gold OA* refers to the practice of publishing scholarly articles in Open Access journals where the articles are freely available to readers without any subscription or paywall barriers. Articles are typically published under a Creative Commons license, which allows for unrestricted distribution and reuse. Authors may be required to pay article processing charges (APCs) to cover the costs of publication.

*Hybrid OA* involves the publication of individual articles in subscription-based journals, but with the option for authors to pay a fee to make their articles openly accessible. This model allows journals to retain subscription revenue while giving authors the option to publish openly.

*Green OA* involves the self-archiving or deposit of scholarly articles in repositories or platforms after publication in subscription-based journals. Articles are made openly accessible through institutional repositories, subject repositories, or preprint servers, providing an additional avenue of access beyond the journal's paywall.

*Bronze OA* includes articles that are openly accessible on a publisher's website without an explicit open license. Some publishers choose to make selected articles freely available even within subscription-based journals. Alternatively, publishers may designate specific journals or sections within journals where articles are accessible without subscription. In some cases, publishers may impose an embargo period during which an article remains behind a paywall. After this period, the article becomes unrestricted access, a practice known as delayed open access.

*Article-level metrics*

The article-level metrics, such as those provided by the Altmetric.com database, offer a different perspective on an article's reach, influence, and impact across different platforms and communities, contributing to a more comprehensive understanding of its importance in both academic and broader contexts. Below there is a description of these article-level metrics.

*Policy citations*: Track the number of times an article has been cited or referenced in policy documents, such as government reports, legislative briefs, or international guidelines. This metric highlights the article's influence on policy-making processes and its contribution to shaping real-world decisions, frameworks, or regulations. Quantifying policy engagement, demonstrates the societal relevance of research beyond academia, particularly in informing evidence-based practices or addressing global challenges.

*Scientific citations*: This fundamental metric tracks the number of times an article has been cited in other scholarly articles indexed in the Dimensions database, indicating its influence and contribution to the scholarly literature in its field.

*Mendeley readers*: Mendeley is a reference management tool used by researchers to organize and share research articles. This metric counts the number of Mendeley users who have added the article to their library, providing insight into the popularity and relevance of the article within the academic community.

*News mentions* count the number of times an article has been mentioned or referenced in news articles, indicating its reach and influence beyond academic circles.

*Blog mentions* measure the number of times an article has been referenced or discussed in blog posts, reflecting its impact on both academic and non-academic online discourse.

*Patent citations* measure the number of times an article has been cited in patents, highlighting its relevance and potential application to innovation and technological development.

*X/Twitter mentions* quantify the number of times an article has been referenced, shared, or discussed on the most widely used social media platform. It reflects the visibility and impact of the article within the X/Twitter community, providing insight into its reception, dissemination, and engagement on this popular social media platform.

*Facebook mentions* count the number of times an article was mentioned or shared on Facebook, indicating its popularity and visibility on this widely used social media platform.

*Wikipedia mentions* track the number of times an article has been referenced or cited on Wikipedia pages, indicating its importance and influence in shaping knowledge and information on the web.

*Video mentions* measure the number of times an article has been referenced or discussed on YouTube, such as in lectures, presentations, or online educational content, demonstrating its impact beyond traditional written media.

*Methods*

An analysis of variance (ANOVA) was used to gain insight into the qualitative factors influencing policy citations. This statistical analysis allowed detect the different variables in explaining variations in policy citations. Specifically, factors such as journal rating, journal quartile, open access type, funding, and publication year were examined.

A logarithmic transformation was applied to account for the typical skewness observed in citation and mention counts. This transformation aimed to normalize the distribution of these counts, which tend to be highly skewed towards lower values. The logarithmic transformation involved taking the natural logarithm (base e) of the variable, with an additional shift of one unit.

A regression model was also used to predict log policy citations based on the selected predictors. The regression analysis made it possible to assess the relationship between the predictor variables and the outcome variable (log policy citations). The signs of the coefficients in the regression model indicated the direction of these relationships. It is important to note, however, that correlation does not imply causation. While the regression model captured associations between variables, it did not establish causal relationships between them.

To interpret the results of the regression analysis, the focus was placed on the standardized coefficients. These coefficients represented the effect size of each predictor variable in units of standard deviation. By examining the standardized coefficients, the relative importance of each predictor in explaining variation in log policy citations could be assessed. This approach facilitated the identification of the most influential factors contributing to the citation in policy documents.

In contrast to the approach taken by Dorta-González et al. (2024), which focused only on journals containing the term 'policy' in their titles, it was decided to expand the database to include all journals categorised as policy-related by the APSA expert group. This expansion was driven by the aim of capturing a broader and more representative sample of policy-related research. By including a wider range of journals, the interdisciplinary nature of policy research can be more fully reflected, ensuring that the analysis covers a wider range of scholarship within the field. In addition, an analysis of APSA journal ratings was included as an indicator of journal quality, alongside two different measures of journal impact. This inclusion was intended to provide a more holistic assessment of the influence and importance of policy research. APSA journal ratings provide valuable insights into the perceived quality and relevance of a journal in the academic sphere, complementing quantitative metrics with qualitative assessments from experts in the field.

Finally, unlike Dorta-González et al. (2024), who aggregated mentions across different social media platforms, we disaggregated mentions by platform and introduced additional categories such as patent and video mentions. This methodology allows for a more granular and nuanced examination of the dissemination and impact of policy research across different channels. By examining mentions separately on different platforms and introducing new categories, a deeper understanding can be gained of how policy research is disseminated and engaged with by different audiences, including academics, policymakers, and the public.

## 4. Results

Table 1 provides an overview of the qualitative data collected on journal articles in the sample, allowing readers to understand trends in journal selection, open-access publishing, funding, and publication over time. The journal expert rating categorizes journals into various levels. The frequencies indicate how many articles were published in journals of each rating. A* and A-rated journals appear to be the most popular choice for publication in policy research, accounting for 45.7% of articles. B-rated journals follow closely with 32.6%, while C-rated journals have the lowest frequency with 21.7%. The SJR best quartiles are used to rank journals according to their impact. Most articles analysed are published in Q1 journals, which account for 80% of the total. However, it should be noted that journals are usually assigned to more than one disciplinary category and that the quartile of a journal often varies depending on the category, so when analysing the best quartile, the data are biased towards the top quartiles.

OA status indicates whether articles are published as Open Access (OA) or not. A considerable proportion, 38.3%, of articles are published as OA, while the majority, 61.7%, are not. For open access articles, OA type indicates the typology of the open access model used. Most OA articles fall into the closed category (61.7%). Hybrid and green OA models have similar frequencies, around 14.0% and 14.8%, respectively. Gold and bronze OA models have a lower frequency, at 2.6% and 6.9%, respectively.

Funding shows the distribution of funded articles being considered. A significant proportion, 21.3%, of articles are funded, while the majority, 78.7%, are not. Finally, the distribution over the years is constant, with around 8% to 14% of articles published each year.

Table 2 presents descriptive statistics for several quantitative variables related to policy research, derived from a dataset of N=149,557 observations. Policy citations range from 0 to 119, with a mean of 0.21 and a standard deviation of 1.14, indicating low citation rates on average, but with significant variability across articles. SJR scores range from 0.10 to 8.35, with a mean of 1.23 and a standard deviation of 1.24, indicating varying levels of journal prestige. Impact scores range from 0.05 to 15.30, with a mean of 3.17 and a standard deviation of 2.33, indicating varying degrees of journal impact. Other variables, including news and blog mentions, patent citations, X and Facebook mentions, Wikipedia and video mentions, Mendeley readers, and scholarly citations, show similar patterns of wide-ranging values, highlighting the diversity across platforms and sources. Overall, these descriptive statistics highlight the breadth and variability within the quantitative measures of policy-related research analysed in this study.

Table 3 shows the results of an ANOVA (analysis of variance) on log policy citations. The logarithmic transformation is used to normalize the distribution, as the counts are typically skewed towards lower values (see Table 2). This ANOVA describes the qualitative factors influencing policy citations, highlighting the importance of journal rating, journal quartile, OA type, funding, and publication year. The different journal ratings have significant effects on policy citations. Higher-rated journals tend to have higher policy citations. The coefficients are all positive and statistically significant ($p < 0.01$) concerning the reference rating C, with A* journals having the highest coefficient (0.034), followed by A and B. Additionally, journal quartiles also have a significant effect on policy citations. Top quartiles are associated with higher policy citations. Q1 has the highest coefficient (0.022), followed by Q2 and Q3. Only Q3 has a p-value above 0.01, indicating a less significant effect compared to the reference quartile Q4.

Compared with closed access, the different types of OA are associated with significantly higher citations in policy documents. Articles published under gold and hybrid open access models have the highest coefficients and are highly significant ($p < 0.0001$), indicating higher policy citations compared to other types. On the other hand, articles without funding receive significantly fewer policy citations than those with funding. Finally, the year of publication also has a significant effect on policy citations. There is a decreasing trend in policy citations over the years, with earlier years having higher coefficients. The coefficient decreases from 0.083 in 2014 to 0.012 in 2022, indicating a decreasing effect over time because they have less time to be cited.

In the context of building a statistical model with log policy citations as the dependent variable, the correlations shown in Table 4 provide insights into potential predictor variables and their relationships with the dependent variable. Let us focus on the correlations that are most relevant to the model. Log scientific citations (0.37) and log Mendeley readers (0.34) have the strongest positive correlations with log policy citations. This suggests that papers with many readers on Mendeley and high citation counts in the scientific literature are also likely to be cited frequently in the policy domain. These metrics may be strong predictors in the model. Journal impact (cites per paper) and journal prestige score (SJR) have positive correlations with citations in policies (0.20 and 0.14 respectively). Although not the strongest, they suggest that papers published in high-prestige journals with higher citation rates may also receive more policy citations. These could also be included as predictors. Log blog mentions (0.16) and news mentions (0.15) have also a weak positive correlation. Including these along with other social media mentions might help attract broader public and policy attention, but their contributions might be weaker.

The highest correlations between the independent variables correspond to Mendeley readers with scientific citations (0.81) and journal prestige score with journal impact (0.77). However, multicollinearity problems in the model can be ruled out. The negative correlation (-0.29) in the article age row suggests that older articles may receive fewer mentions on X/Twitter. However, the positive correlation with log scientific citations (0.39) and log Mendeley readers (0.30) suggests a time lag in academic recognition. Including article age as a predictor may be useful to account for this.

The key elements of a regression model predicting the number of citations (log-transformed) that policy papers receive in policy documents are given in Table 5. The adjusted goodness-of-fit statistic $R^2$ is 0.157 and represents the proportion of variance in log policy citations explained by the model after considering the number of predictors (19 in this case). Although this is not a high value, it suggests that the model captures some of the systematic variation in citations. In the analysis of variance (ANOVA), the highly significant F-statistic (p-value < 0.0001) indicates that the model statistically explains a non-zero proportion of the variance in log policy citations.

Table 6 shows the estimated coefficients, standard errors, t-statistics, p-values, and other statistics for the parameters of the regression model predicting log policy citations. The p-value significance codes help to identify statistically significant relationships between the predictors and log policy citations. The signs of the coefficients indicate the direction of the relationship (e.g. a positive coefficient for journal impact means that higher impact is associated with more policy citations). It is important to note that the model captures associations, not necessarily causal relationships. Standardized coefficients represent the effect size of each predictor in units of standard deviation. For example, a one standard deviation increase in journal impact is associated with a 0.089 increase in log policy citations (controlling for other factors).

About journal characteristics, journal expert rating and journal prestige score (SJR) have significant negative coefficients. This suggests that papers published in high-rated journals with higher prestige are less likely to be cited in policies. Note that being in the top quartiles (SJR) have also significant negative coefficient. However, journal impact (cites per paper) has a significant positive coefficient. This suggests that papers published in high-impact journals are more likely to be cited in policy documents. This apparent contradiction between the journal prestige score and the citations per document is due to the different methodologies used in both impact indicators at the journal level. While in the impact factor type indicator (citations per document) all citations have the same value, in the SJR prestige indicator the weight of each citation is proportional to the impact of the citing journal.

Looking at article characteristics, most of the open access (OA) types show significant and positive effects compared to the closed access modality. Specifically, hybrid OA status is associated with the highest effect size, followed by gold OA with a slightly smaller effect size. Notably, the effect of green OA is not statistically significant. These findings suggest that hybrid and gold OA modalities may tend to encourage policy citations, albeit to varying degrees, while closed access may potentially discourage such citations.

A small but significant positive coefficient indicates that policy citations may increase with article age. Similarly, policy citations may also increase for funded articles, although in this case, the effect is less significant (p-value 0.04).

Online mentions and citations have a significant impact on policy visibility and impact. Academic platforms such as Dimensions and Mendeley show the strongest effect, highlighting the importance of scientific recognition. Popular media, with news and blog mentions, also play a role. Even Facebook mentions have a positive, albeit smaller, impact.

The results above revealed a notable correlation between Mendeley readership and policy citations (see Table 4), which initially led to an investigation into whether this association might be accounted for by other underlying characteristics. However, when all variables were included in our multivariate model, we found that although both journal impact and year of publication contributed to the observed relationship — with older publications and those in high-impact journals naturally gaining greater visibility — the strength of the association between Mendeley usage and policy citations remained robust. This suggests that, while part of the correlation can indeed be attributed to these confounding factors, engagement on Mendeley appears to independently increase the likelihood of research being referenced in policy documents.

Interestingly, X/Twitter mentions show a negative correlation with citations in policy documents, indicating that such platforms may not significantly contribute to policy-relevant discourse. This finding suggests X/Twitter does not serve as an effective channel for discussions that directly inform evidence-based policymaking. The results align with existing critiques of social media's limited capacity to facilitate substantive policy engagement, despite its broader sociopolitical visibility.

Moreover, video mentions demonstrate no statistically significant relationship with citations in policy documents. This highlights the fundamental distinction between formal policy documentation — which prioritises rigorous, evidence-based frameworks — and more accessible media formats. While video platforms may enhance public awareness, their influence on professional policy citation practices appears negligible.

Given the magnitude of the effects, the variables can be compared with each other and with a reference variable, namely the effect of time on the accumulation of citations in policy documents (article age). They are then classified into five groups according to their influence:

1) Strong effect: Scientific citations have the strongest positive effect (standardized coefficient 0.247). This effect is more than eight times higher than the time effect, highlighting the central role of scientific citations in shaping policy decisions.

2) Medium effect: Articles published in journals with higher citation impact (citations per document in 3 years) tend to be cited more often in policy documents. Similarly, articles with higher readership on Mendeley show comparable effects (0.089 and 0.086 respectively). In both cases, the effect is about three times as large as the effect associated with ageing over the years.

3) Small effect: News mentions and blog engagement both show significant positive effect sizes (0.057 and 0.054 respectively, i.e. twice and 1.9 times the effect of ageing). In addition, hybrid access (OA with publisher fees) shows a positive effect (0.049, or 1.7 times the effect of ageing) on policy citations.

4) Small negative effect: Mentions on X/Twitter show a negative effect (-0.040, or 1.4 times the effect of ageing in absolute terms).

5) Negligible negative effect: Interestingly, a higher prestige score derived from the SJR correlates with a small negative effect on policy citations (-0.018). Similarly, publication in a top-quartile journal based on SJR scores also has a negative effect (-0.028). Furthermore, a higher expert rating for a journal shows a modest negative effect (-0.014). However, all these effects are smaller in absolute terms than simple ageing.

## 5. Discussion

The results underscore the varying degrees of influence that different factors have on the citation of articles within policy documents. For the sake of comparison, the age variable is used as a benchmark. The effect of ageing, or the passage of time, reflects the natural process of older articles accumulating citations over time. Given that the variable analysed is the number of policy citations, it is naturally expected to increase over time.

The model variables of this study show improved significance and explanatory power compared to Dorta-González et al. (2024). This is because the present study expands the database to include all journals classified as policy-related by the APSA expert panel, instead of those with the word policy in their title, ensuring a more comprehensive and representative sample. Furthermore, APSA journal ratings and two impact measures are used to assess the influence of the journal, while mentions are disaggregated by platform, and new categories are added for more detailed analysis. The main difference in the results was found in the effect of the scientific citations, with an increase in statistical significance from 0.04 to below 0.0001 and a substantial increase in the standardized coefficient from 0.009 to 0.247, indicating a stronger relationship between scientific and policy citations.

*Strong effect: Scientific references*

The substantial difference in effect sizes between scientific citations and ageing underscores the critical role of scientific literature in informing policy decisions. The fact that the effect of scientific citations is more than eight times greater than that of ageing highlights the capital importance of evidence-based research in shaping policies and interventions that address societal challenges effectively.

Scientific citations represent direct references made by researchers to previously published scientific work. These citations indicate the influence and relevance of a particular study within the scientific community. Scientific citations are crucial indicators of the quality, significance, and trustworthiness of research findings. In the context of policy decisions, citations from scientific literature carry substantial weight as they signify evidence-based support for policies or interventions. Therefore, a higher number of scientific citations suggests a stronger foundation of empirical evidence supporting a given policy, which can significantly influence decision-makers.

*Moderate impact: Journal influence and Mendeley readership*

The medium positive effect, approximately three times the impact of ageing, observed for both journal impact and Mendeley readership indicates that factors related to the influence and engagement with scholarly articles significantly influence their policy impact. Policymakers are more likely to consider findings from reputable journals with high-impact factors and articles with active engagement on platforms like Mendeley when making informed decisions, highlighting the importance of these factors in bridging the gap between research and policy.

These results can be justified by considering the dynamics of scholarly communication and the role of these factors in influencing policy impact. Firstly, the impact factor of a journal serves as a widely recognized metric for assessing the influence of academic publications within a particular field. Articles published in journals with higher impact factors are often perceived as more reputable, rigorous, and impactful within the scientific community. As a result, policymakers may prioritize findings from these journals when formulating policies, as they are more likely to represent high-quality research supported by robust evidence and rigorous peer review processes. Therefore, the medium positive effect observed for articles from journals with higher impact factors suggests that the influence of the journal contributes significantly to its policy impact.

Secondly, Mendeley's readership reflects the extent to which an article is being actively engaged with and discussed within the academic community. Mendeley is a platform commonly used by researchers to organize and share scholarly articles, annotate, and highlight key findings, and engage in discussions with peers. Higher readership on Mendeley indicates increased interest and engagement with the article's content, suggesting its relevance and significance within the scholarly community. Policymakers may take note of articles with higher Mendeley readership as they may signify emerging trends, innovative approaches, or critical insights relevant to policy formulation. Therefore, the medium positive effect observed for articles with greater readership on Mendeley suggests that active engagement and discussion within the academic community contribute to the article's policy impact.

*Minor impact: News coverage, Blog engagement, and hybrid accessibility*

News mentions and blog engagement have a positive effect on policy citations, about twice and 1.9 times the impact of ageing, respectively. Hybrid access also has a positive influence, with 1.7 times the effect of ageing. These significative effects observed for news mentions, blog engagement, and hybrid access indicate that broader dissemination and accessibility can moderately increase an article's policy influence. By reaching a wider audience and ensuring greater accessibility to research findings, these factors contribute to bridging the gap between research and policy, thereby facilitating evidence-informed decision-making processes.

Firstly, news and blog engagement signify broader public exposure and engagement with research findings outside of traditional academic circles. News and blogs function as a spark for initial public awareness and participation. Early exposure sparks the interest of a wider audience and lays the groundwork for later interactions in the political, societal, and academic spheres (see Ortega 2021). When research findings are covered in news articles or discussed on prominent blogs, they reach a wider audience beyond academia, including policymakers, journalists, and the public. Policymakers may take note of research findings that receive significant media attention, as they may reflect topics of public interest or emerging issues with potential policy implications.

Secondly, hybrid access, which combines open access with publisher fees, enhances the accessibility of research findings by providing both subscription-based and freely accessible versions of articles. This accessibility can facilitate greater dissemination and uptake of research findings by policymakers, practitioners, and other stakeholders involved in policy decision-making. The small positive effect observed for hybrid access suggests that increasing the accessibility of research articles through innovative publishing models can modestly enhance their policy impact by ensuring that relevant stakeholders have access to the latest evidence and insights.

It was observed that certain accessibility factors can have a significant impact on policy citation rates in comparison to closed access, where publication in paywalled journals can function as a barrier to policymakers. Research made available through open access channels, particularly hybrid models, has a greater potential to bridge the gap between policy considerations and academic evidence. These results confirm others previously obtained in the case of academic citations (see Dorta-González & Dorta-González 2023a, b). However, no significant relationship was found between policy citations and accessibility through institutional or thematic open repositories (green OA). This implies that despite the availability of research through green OA channels, policymakers were not significantly influenced or motivated to cite these sources in their policy-making processes. This finding is interesting because it contrasts with the positive correlation observed for research made available through other open-access channels, suggesting that the mere accessibility of research through green OA repositories might not necessarily translate into increased policy citations.

*Minor adverse impact: Social media references*

Social media mentions reflect the modern environment where digital platforms function as dynamic forums for the exchange of ideas. These exchanges take place in real time, speeding up the incorporation of research findings into wider public discourses and potentially influencing policy deliberations in the process (see Gong et al., 2023). However, social media mentions, particularly on platforms like X/Twitter, have a small negative effect, with 1.4 times the impact of ageing in absolute terms. This could suggest that the informal nature of social media discourse might detract from the perceived credibility of scientific findings.

The justification for the small negative effect observed for social media mentions, particularly on platforms like X/Twitter, lies in the informal nature of social media discourse, which may detract from the perceived credibility of scientific findings. Firstly, social media platforms like X/Twitter are characterized by rapid information dissemination and limited character counts, which can lead to oversimplification or misinterpretation of complex scientific concepts and findings. Unlike traditional academic channels, where research findings undergo peer review and rigorous scrutiny before publication, social media allows for the instantaneous sharing of information without the same level of quality control. As a result, scientific findings shared on social media platforms may be prone to misrepresentation, misinformation, or selective reporting, which could undermine their credibility among policymakers and other stakeholders.

Secondly, the brevity and informal tone of social media interactions may not adequately convey the nuance and complexity of scientific research. Policymakers may perceive information shared on social media platforms as less authoritative or reliable compared to peer-reviewed academic publications or reputable news sources. Consequently, social media mentions of research findings may not carry the same weight or influence in policy decision-making processes as citations from more traditional and established sources.

*Negligible adverse impact: Journal reputation and expert evaluations*

Surprisingly, higher prestige scores from the SJR and publications in top quartile journals based on SJR scores correlate with a slight negative effect on policy citations. A higher expert rating also shows a small negative effect. However, these effects are smaller than the impact of ageing, indicating that while journal prestige and expert opinions are important, they may not always align with policy relevance.

The small negative effect on policy citations for journals with higher prestige scores from the SJR and those in the top quartile based on SJR scores may be attributed to several factors. While these journals are esteemed within the academic community and often attract high-quality research, their focus may lean more toward theoretical or specialized topics that may not directly translate to immediate policy implications. Policymakers may prioritize research from journals that address pressing societal challenges or provide actionable insights for policy formulation, even if they are not considered prestigious within academic circles. Therefore, the small negative effect observed for journal prestige scores suggests that scholarly recognition alone may not guarantee policy impact.

Similarly, the small negative effect observed for higher APSA journal ratings indicates that while expertise and authority are valuable attributes in academic discourse, they may not always translate to increased policy citations. Policymakers may value diverse perspectives and evidence from multiple sources when making decisions, rather than relying solely on the opinions of individual experts or expert panels. Additionally, the expertise of researchers or reviewers may not always align with the specific needs or priorities of policymakers, leading to a disconnect between APSA journal ratings and policy relevance.

*Apparent contradiction in the sign of effects at the journal level*

The findings reveal a surprisingly significant negative association between journal quality, as assessed by the journal's expert rating and prestige score, and policy citations, suggesting that articles from highly rated journals with greater prestige are less likely to be cited in policy documents. However, the impact factor type metric, specifically citations per paper, shows a positive correlation with policy citations, suggesting an inverse relationship between journal prestige and impact in the policy domain.

This observed variation in model coefficients for journal-level variables may be due to the inherent characteristics of the metrics themselves. Specifically, when looking at articles from journals with similar citation averages, articles published in journals that are frequently cited by other prestigious journals or are highly regarded by experts tend to be more theoretical in content and less practical in application. This theoretical orientation may lead to fewer citations in policy-related contexts. This trend suggests that while such articles have considerable influence within academic circles, their direct impact on policymaking may be limited (see Dorta-González et al., 2024).

On the other hand, in academic circles, research published in highly reputed journals may not be frequently cited in policy documents, despite their academic influence. These prestigious publications often have a dual nature: they are widely cited in academic communities, but less frequently referenced in policy contexts. This divergence can be attributed to their different priorities. Prestigious journals often prioritize research that advances theoretical understanding in a field, which does not necessarily translate into practical solutions to real-world problems. Conversely, policy documents prioritize actionable research that offers immediate applications to policy challenges. Therefore, the different objectives of prestigious academic journals and policy documents are the driving force behind the apparent disparity in citation patterns.

*Limitations of the study*

Policy documents may not comprehensively capture all instances where research informs policy decisions. In addition, reliance on citations within these documents may introduce potential biases, as certain policy areas may be more likely to cite research than others. Given the range of methods available for assessing the impact of policy research, the focus was placed on citations within policy documents for several reasons. First, this approach engages directly with the policy sphere and facilitates the identification of research articles that have a tangible impact on policy development. As a result, it provides a clearer understanding of the practical application of scientific evidence. Secondly, policy citations provide measurable evidence of impact, enabling statistical analysis. This helps to identify trends and patterns in the use of research by policymakers. However, the inherent limitations of this approach are recognized.

In sum, policy citations are influenced by a complex interplay between scientific research and its impact on society through policy discourse. Our findings deepen the understanding of this dynamic and provide valuable insights for researchers, policymakers, and stakeholders seeking to maximize the practical application of scientific knowledge. These insights also have practical implications for those seeking to broaden and enhance their contributions to shaping public policy agendas.

Table 3 revealed pronounced non-linear relationships between policy citations and certain key variables, such as publication year and journal quartile or expert rating classes. This suggests that when these factors are rendered as continuous variables, as in Table 6, their true predictive capacity may be understated. In essence, the continuous specification may not fully capture the nuanced, discrete changes observed in different publication cohorts or quality strata, potentially obscuring important variations in how these dimensions influence policy citations. A more detailed examination, perhaps through the retention of their categorical nature or the use of non-linear modelling techniques, would likely provide a clearer picture of the specific mechanisms through which these variables exert their influence on the integration of scientific research into policy-making processes.

*Considerations about causal inference and bidirectional relationships*

It is important to clarify that, although our analysis reveals significant correlations between scholarly citations, journal impact metrics, and policy document citations, these associations should not be construed as evidence of direct causality. In reality, these relationships may be substantially influenced by confounding factors or even reverse causalities. For example, it is conceivable that research articles receiving attention in policy documents might subsequently experience an increase in scholarly citations, rather than scholarly citations directly prompting policy document references. Consequently, while our findings suggest that highly cited and widely read papers are more frequently referenced in policy contexts, we caution against drawing definitive causal inferences based solely on these correlations. Future work employing longitudinal or quasi-experimental designs would be instrumental in disentangling these complex interdependencies and validating the directional effects observed in this study.

Table 1. Frequencies for qualitative data (N=149,557)

| Variables | Categories | Num articles | % |
|---|---|---|---|
| Journal expert rating | A* (higher) | 24,849 | 16.6 |
| | A | 43,560 | 29.1 |
| | B | 48,762 | 32.6 |
| | C (lower) | 32,386 | 21.7 |
| Journal best quartile SJR | Q1 (higher) | 119,718 | 80.0 |
| | Q2 | 22,387 | 15.0 |
| | Q3 | 6,046 | 4.0 |
| | Q4 (lower) | 1,406 | 1.0 |
| OA status | No | 92,262 | 61.7 |
| | Yes | 57,295 | 38.3 |
| OA type | closed | 92,261 | 61.7 |
| | hybrid | 20,938 | 14.0 |
| | green | 22,202 | 14.8 |
| | gold | 3,903 | 2.6 |
| | bronze | 10,253 | 6.9 |
| Funding | No | 117,764 | 78.7 |
| | Yes | 31,793 | 21.3 |
| Publication year | 2014 | 12,565 | 8.4 |
| | 2015 | 12,064 | 8.1 |
| | 2016 | 20,959 | 14.0 |
| | 2017 | 14,454 | 9.7 |
| | 2018 | 14,236 | 9.5 |
| | 2019 | 14,421 | 9.6 |
| | 2020 | 15,870 | 10.6 |
| | 2021 | 16,018 | 10.7 |
| | 2022 | 15,146 | 10.1 |
| | 2023 | 13,824 | 9.2 |

Table 2. Descriptive statistics for quantitative data (N=149,557)

| Variables | Min | Max | Mean | SD |
|---|---|---|---|---|
| Policy citations | 0 | 119 | 0.209 | 1.142 |
| Journal prestige score SJR | 0.101 | 8.346 | 1.233 | 1.237 |
| Journal impact (cites per doc 3 years) | 0.053 | 15.297 | 3.168 | 2.331 |
| News mentions | 0 | 386 | 0.545 | 3.858 |
| Blog mentions | 0 | 122 | 0.212 | 0.820 |
| Patent citations | 0 | 33 | 0.002 | 0.109 |
| X mentions | 0 | 16316 | 11.971 | 77.324 |
| Facebook mentions | 0 | 133 | 0.139 | 0.672 |
| Wikipedia mentions | 0 | 47 | 0.156 | 0.840 |
| Video mentions | 0 | 12 | 0.002 | 0.065 |
| Mendeley readers | 0 | 9939 | 37.615 | 78.733 |
| Scientific citations | 0 | 4669 | 18.333 | 58.373 |

Table 3. ANOVA for log Policy citations

| Source | Coef. | Stand. error | t | Pr > \|t\| | Lower limit (95%) | Upper limit (95%) | p-values codes |
|---|---|---|---|---|---|---|---|
| Interception | -0.029 | 0.004 | -7.212 | **<0.0001** | -0.036 | -0.021 | *** |
| Journal expert rating-A* | 0.034 | 0.001 | 27.066 | **<0.0001** | 0.032 | 0.037 | *** |
| Journal expert rating-A | 0.013 | 0.001 | 11.289 | **<0.0001** | 0.010 | 0.015 | *** |
| Journal expert rating-B | 0.004 | 0.001 | 3.442 | **0.001** | 0.002 | 0.006 | *** |
| Journal expert rating-C | 0.000 | 0.000 | | | | | |
| Journal best quartile SJR-Q1 | 0.022 | 0.004 | 5.877 | **<0.0001** | 0.015 | 0.030 | *** |
| Journal best quartile SJR-Q2 | 0.016 | 0.004 | 4.241 | **<0.0001** | 0.009 | 0.024 | *** |
| Journal best quartile SJR-Q3 | 0.009 | 0.004 | 2.216 | **0.027** | 0.001 | 0.017 | * |
| Journal best quartile SJR-Q4 | 0.000 | 0.000 | | | | | |
| OA type-hybrid | 0.040 | 0.001 | 34.992 | **<0.0001** | 0.038 | 0.043 | *** |
| OA type-green | 0.018 | 0.001 | 17.041 | **<0.0001** | 0.016 | 0.020 | *** |
| OA type-gold | 0.043 | 0.002 | 18.423 | **<0.0001** | 0.039 | 0.048 | *** |
| OA type-bronze | 0.012 | 0.001 | 8.184 | **<0.0001** | 0.009 | 0.015 | *** |
| OA type-closed | 0.000 | 0.000 | | | | | |
| Funding-No | -0.022 | 0.001 | -24.512 | **<0.0001** | -0.024 | -0.021 | *** |
| Funding-Yes | 0.000 | 0.000 | | | | | |
| Publication year-2014 | 0.083 | 0.002 | 47.309 | **<0.0001** | 0.080 | 0.087 | *** |
| Publication year-2015 | 0.072 | 0.002 | 40.585 | **<0.0001** | 0.068 | 0.075 | *** |
| Publication year-2016 | 0.077 | 0.002 | 48.751 | **<0.0001** | 0.073 | 0.080 | *** |
| Publication year-2017 | 0.064 | 0.002 | 38.108 | **<0.0001** | 0.061 | 0.068 | *** |
| Publication year-2018 | 0.060 | 0.002 | 35.629 | **<0.0001** | 0.057 | 0.064 | *** |
| Publication year-2019 | 0.047 | 0.002 | 28.271 | **<0.0001** | 0.044 | 0.051 | *** |
| Publication year-2020 | 0.037 | 0.002 | 22.607 | **<0.0001** | 0.034 | 0.040 | *** |
| Publication year-2021 | 0.026 | 0.002 | 15.931 | **<0.0001** | 0.023 | 0.029 | *** |
| Publication year-2022 | 0.012 | 0.002 | 7.332 | **<0.0001** | 0.009 | 0.015 | *** |
| Publication year-2023 | 0.000 | 0.000 | | | | | |

Signification codes: *** very highly significant; ** highly significant; * significant; 0 < *** < 0.001 < ** < 0.01 < * < 0.05

Table 4. Correlations

| | Journal prestige score SJR | Journal impact (cites per doc 3 years) | Article age (years) | log News mentions | log Blog mentions | log Patent citations | log X mentions | log Facebook mentions | log Wikipedia mentions | log Video mentions | log Mendeley readers | log Scientific citations | **log Policy citations** |
|---|---|---|---|---|---|---|---|---|---|---|---|---|---|
| Journal prestige score SJR | **1** | 0.77 | 0.08 | 0.10 | 0.18 | 0.02 | 0.15 | 0.00 | 0.09 | 0.01 | 0.23 | 0.27 | **0.14** |
| Journal impact (cites per doc 3 years) | 0.77 | **1** | 0.02 | 0.09 | 0.17 | 0.02 | 0.13 | 0.00 | 0.02 | 0.02 | 0.41 | 0.36 | **0.20** |
| Article age (years) | 0.08 | 0.02 | **1** | 0.01 | 0.02 | 0.01 | -0.29 | 0.13 | 0.09 | 0.00 | 0.30 | 0.39 | **0.15** |
| log News mentions | 0.10 | 0.09 | 0.01 | **1** | 0.38 | 0.01 | 0.24 | 0.09 | 0.13 | 0.09 | 0.20 | 0.23 | **0.15** |
| log Blog mentions | 0.18 | 0.17 | 0.02 | 0.38 | **1** | 0.01 | 0.24 | 0.10 | 0.10 | 0.08 | 0.22 | 0.25 | **0.16** |
| log Patent citations | 0.02 | 0.02 | 0.01 | 0.01 | 0.01 | **1** | -0.01 | 0.00 | 0.00 | 0.01 | 0.02 | 0.02 | **0.03** |
| log X mentions | 0.15 | 0.13 | -0.29 | 0.24 | 0.24 | -0.01 | **1** | 0.14 | 0.04 | 0.05 | 0.20 | 0.16 | **0.05** |
| log Facebook mentions | 0.00 | 0.00 | 0.13 | 0.09 | 0.10 | 0.00 | 0.14 | **1** | 0.03 | 0.04 | 0.10 | 0.11 | **0.06** |
| log Wikipedia mentions | 0.09 | 0.02 | 0.09 | 0.13 | 0.10 | 0.00 | 0.04 | 0.03 | **1** | 0.04 | 0.07 | 0.12 | **0.06** |
| log Video mentions | 0.01 | 0.02 | 0.00 | 0.09 | 0.08 | 0.01 | 0.05 | 0.04 | 0.04 | **1** | 0.04 | 0.04 | **0.03** |
| log Mendeley readers | 0.23 | 0.41 | 0.30 | 0.20 | 0.22 | 0.02 | 0.20 | 0.10 | 0.07 | 0.04 | **1** | 0.81 | **0.34** |
| log Scientific citations | 0.27 | 0.36 | 0.39 | 0.23 | 0.25 | 0.02 | 0.16 | 0.11 | 0.12 | 0.04 | 0.81 | **1** | **0.37** |
| **log Policy citations** | **0.14** | **0.20** | **0.15** | **0.15** | **0.16** | **0.03** | **0.05** | **0.06** | **0.06** | **0.03** | **0.34** | **0.37** | **1** |

Note: The log refers to the natural logarithm in base e of the variable shifted by one unit.

Table 5. Analysis of variance (log Policy citations)

| Source | GL | Sum of squares | Sum of squares | F | Pr > F |
|---|---|---|---|---|---|
| Model | 19 | 475.868 | 25.046 | 1471.214 | **<0.0001** |
| Error | 149537 | 2545.692 | 0.017 | | |
| Total corrected | 149556 | 3021.561 | | | |

Note: Calculated against the model Y=Mean(Y). Adjusted R² = 0.157.

Table 6. Model parameters (log Policy citations)

| Source | Coef. | Stand. error | t | Pr > \|t\| | Lower limit (95%) | Upper limit (95%) | p-values codes | Stand. coef. |
|---|---|---|---|---|---|---|---|---|
| **Journal level metrics** | | | | | | | | |
| Journal expert rating (higher better) | -0.002 | 0.000 | -4.387 | **<0.0001** | -0.003 | -0.001 | *** | -0.014 |
| Journal prestige score SJR | -0.002 | 0.000 | -4.252 | **<0.0001** | -0.003 | -0.001 | *** | -0.018 |
| Journal best quartile SJR (top better) | -0.007 | 0.001 | -10.706 | **<0.0001** | -0.008 | -0.006 | *** | -0.028 |
| Journal impact score (cites per doc 3 years) | 0.005 | 0.000 | 20.452 | **<0.0001** | 0.005 | 0.006 | *** | 0.089 |
| **Article attributes** | | | | | | | | |
| Article age (years) | 0.001 | 0.000 | 9.555 | **<0.0001** | 0.001 | 0.002 | *** | 0.029 |
| Funding (Yes=1, No=0) | 0.002 | 0.001 | 2.057 | **0.040** | 0.000 | 0.004 | * | 0.005 |
| OA (relative to closed) | | | | | | | | |
|    OA hybrid | 0.020 | 0.001 | 18.155 | **<0.0001** | 0.018 | 0.022 | *** | 0.049 |
|    OA green | 0.001 | 0.001 | 1.115 | 0.265 | -0.001 | 0.003 | ° | 0.003 |
|    OA gold | 0.011 | 0.002 | 4.863 | **<0.0001** | 0.006 | 0.015 | *** | 0.012 |
|    OA bronze | 0.006 | 0.001 | 4.623 | **<0.0001** | 0.004 | 0.009 | *** | 0.011 |
| **Article level metrics** | | | | | | | | |
| log News mentions | 0.038 | 0.002 | 21.625 | **<0.0001** | 0.034 | 0.041 | *** | 0.057 |
| log Blog mentions | 0.055 | 0.003 | 20.321 | **<0.0001** | 0.050 | 0.061 | *** | 0.054 |
| log Patent citations | 0.227 | 0.025 | 9.196 | **<0.0001** | 0.179 | 0.275 | *** | 0.022 |
| log X mentions | -0.010 | 0.001 | -14.248 | **<0.0001** | -0.012 | -0.009 | *** | -0.040 |
| log Facebook mentions | 0.012 | 0.003 | 3.865 | **<0.0000** | 0.006 | 0.018 | *** | 0.009 |
| log Wikipedia mentions | 0.008 | 0.003 | 2.791 | **0.005** | 0.002 | 0.013 | ** | 0.007 |
| log Video mentions | 0.036 | 0.022 | 1.641 | 0.101 | -0.007 | 0.079 | ° | 0.004 |
| log Mendeley readers | 0.020 | 0.001 | 19.412 | **<0.0001** | 0.018 | 0.022 | *** | 0.086 |
| log Scientific citations | 0.057 | 0.001 | 56.597 | **<0.0001** | 0.055 | 0.059 | *** | 0.247 |
| Interception | -0.058 | 0.002 | -32.023 | <0.0001 | -0.062 | -0.055 | *** | - |

Signification codes: *** very highly significant; ** highly significant; * significant; . marginally significant; ° not significant; 0 < *** < 0.001 < ** < 0.01 < * < 0.05 < . < 0.1 < ° < 1. The log refers to the natural logarithm in base e of the variable shifted by one unit.